\documentclass[11pt, a4paper]{main}

\usepackage[authoryear, sort&compress, round]{natbib}
\usepackage{graphicx}
\usepackage{subcaption}
\usepackage{tcolorbox}
\usepackage{xspace}
\usepackage{booktabs}
\usepackage{multirow}
\usepackage{amsthm}
\usepackage{tikz}
\usepackage{pifont}
\usepackage{hyperref}
\usepackage{xurl}
\usepackage{enumitem}
\usepackage{listings}
\tcbuselibrary{listings}
\tcbuselibrary{skins}
\usepackage{pdfpages}
\usepackage{algorithm}
\usepackage{algorithmic}
\usepackage{latexsym}
\usepackage[T1]{fontenc}
\usepackage[utf8]{inputenc}

\usepackage{microtype}

\newcommand{\code}[1]{\texttt{#1}}

\AtBeginDocument{%
  }

\lstdefinelanguage{Python}{
    morekeywords={class, def, return, try, except, raise, from},
    keywordstyle=\color{blue},
    stringstyle=\color{green},
    commentstyle=\color{gray},
    morecomment=[l]{\#},
}

\lstdefinelanguage{diff}{
    morecomment=[f][\color{green}]{+},
    morecomment=[f][\color{red}]{-},
    morecomment=[f][\color{blue}]{@@},
}

\lstdefinelanguage{errorlog}{
    morecomment=[f][\color{red}]{E},
    morecomment=[f][\color{magenta}]{?},
}
\newcommand{\tool}{\textsc{SWE-Synth}\xspace}

\renewcommand{\cite}{\citep}
\title{\textcolor{blue!70!black}{\textsc{SWE-Synth}}: Synthesizing Verifiable Bug-Fix Data to Enable Large Language Models in Resolving Real-World Bugs}

\author[1]{Minh V.T. Pham\textsuperscript{*}}
\author[1]{Huy N. Phan\textsuperscript{*}}
\author[3]{Hoang N. Phan}
\author[1]{Cuong Le Chi}
\author[2]{Tien N. Nguyen}
\author[1]{Nghi D. Q. Bui\textsuperscript{$\dagger$}}

\affil[1]{FPT Software AI Center, Viet Nam}
\affil[2]{University of Texas at Dallas, US}
\affil[3]{Nanyang Technological University, Singapore}
\begin{document}

% Applying large language models (LLMs) to autonomously fix software bugs is a promising direction that could reshape the software development landscape. However, their effectiveness is currently limited by the lack of diverse, real-world bug datasets necessary for training. While combining large-scale mining with human curation can produce such datasets, the approach is costly and does not scale well. To overcome this limitation, we propose a novel and scalable synthetic data generation pipeline that uses LLMs to create realistic bugs via targeted code rewriting. Our pipeline also synthesizes valuable intermediate repair steps, enhancing the training signal for accurate bug fixing. Using this approach, we construct {\tool}, a large, diverse, and context-rich dataset of bug-fix pairs that are natural, scalable, automatically verifiable, and include intermediate reasoning steps. LLMs trained on {\tool} learn context-aware repair strategies and achieve accuracy comparable to models trained on manually curated datasets like SWE-Gym, while offering significantly greater scalability and efficiency. We demonstrate the effectiveness of our method on standard benchmarks such as SWE-Bench and BugsInPy.

\begin{abstract}
Large language models (LLMs) are transforming automated program repair (APR) through agent-based approaches that localize bugs, generate patches, and verify fixes. However, the lack of high-quality, scalable training datasets—especially those with verifiable outputs and intermediate reasoning traces—limits progress, particularly for open-source models. In this work, we present \tool, a framework for synthesizing realistic, verifiable, and process-aware bug-fix datasets at the repository level. \tool leverages LLM agents to simulate debugging workflows, producing not only bug–fix pairs but also test cases and structured repair trajectories. Compared to manually curated datasets, our method scales with minimal human effort while preserving contextual richness and correctness. Experiments show that models trained on \tool outperform those trained on real-world datasets by 2.3\% on SWE-Bench Lite. Our results highlight the potential of synthetic, agent-generated data to advance the state of the art in APR and software engineering automation. Our dataset and code are publicly available at~\url{https://github.com/FSoft-AI4Code/SWE-Synth}.
\end{abstract}

\footnotetext[1]{*Co-author with equal contribution}
\footnotetext[2]{$\dagger$Project Lead}
\maketitle

\section{Introduction}

Large language models (LLMs) have made significant strides in automating software engineering (SE) tasks, from code generation~\cite{wei2023magicoder, chen2021evaluating, wang2023codet5+, li2022competition, zhuo2024bigcodebench, bui2023codetf, manh2023vault, to2023functional} to bug fixing~\cite{jimenez2023swe, xia2024agentless} and test synthesis~\cite{chen2022codet, he2025hardtests, li2024large}. Recent efforts have explored the development of LLM-based agents that can autonomously navigate repositories, localize bugs, generate patches, and verify correctness~\cite{yang2024swe, jain2024r2e, liu2024large, phan2024hyperagent, nguyen2024agilecoder}. These agent-based approaches promise to scale software engineering workflows, but they require training data that is both high-quality and task-aligned.

Benchmarks like SWE-Bench~\cite{jimenez2023swe} and SWE-Bench Lite~\cite{yang2024swe} have emerged as standard evaluations for assessing autonomous software engineering (SWE) agents on realistic tasks such as GitHub issue resolution. These benchmarks have revealed a growing performance gap between proprietary and open-source models~\cite{xie2025swe, jaech2024openai}. While commercial agents backed by closed models (e.g., GPT-4, Claude 3.5) achieve strong performance~\cite{anthropic2025claude37}, open models still lag behind, in part due to a lack of high-quality training environments and scalable datasets. Unfortunately, existing datasets often fall short. Mined bug-fix pairs from open-source repositories~\cite{just2014defects4j, herzig2013it} are noisy, lack intermediate reasoning traces, and rarely include test-based verifiability~\cite{yan2024enhancing, yang2024swe}. Even curated benchmarks like SWE-Bench contain gold patches but lack comprehensive, executable environments for training. SWE-Gym~\cite{pan2024training}, for example, partially addresses this by offering environments for a limited number of real issues, but its reliance on human-curated test cases limits its scalability. Synthetic data generation offers a promising, scalable alternative~\cite{long2024llms, wang2024openhands}, yet existing techniques have their own shortcomings. Mutation-based methods~\cite{jia2011analysis} often yield trivial or unrealistic bugs. LLM-based synthetic data generation, while successful in instruction tuning~\cite{peng2023instructiontuninggpt4, li2023self}, is not yet mature for complex SE tasks like bug repair, fault localization, or test-driven development. Furthermore, synthetic data frequently lacks contextual grounding or verification, limiting its usefulness for downstream models~\cite{ding-etal-2023-gpt, pan2024training}.

We believe that progress in LLM-based program repair hinges on synthetic datasets that are not just large, but realistic, verifiable, and process-aware. In particular, we identify four key properties that such datasets should possess:

\begin{itemize}[leftmargin=*]
    \item \textbf{Human-like bugs:} The dataset should feature natural errors that resemble those made by real developers~\cite{ou2024synatra, xie2025swe}.
    \item \textbf{Scalability:} The data generation process must be efficient and capable of scaling to large codebases without human intervention~\cite{jain2024r2e, trabucco2025insta}.
    \item \textbf{Automated verification:} Each buggy example should be paired with test cases that verify both the existence of the bug and the correctness of the fix~\cite{le2022coderl, li2025gift, acecoder, chu2025sftmemorizesrlgeneralizes}.
    \item \textbf{Repair traces:} Intermediate steps—such as localization, fixing, and refinement—should be included to model the developer’s iterative reasoning process~\cite{yang2024swe, stechly2024chain, zhou2024archer}.
\end{itemize}

In this work, we propose \tool, a scalable framework for generating high-fidelity, verifiable synthetic bug-fix datasets using LLM agents that simulate human debugging workflows. Our pipeline produces not only buggy-fix pairs, but also structured intermediate traces and test logs that enable modern learning algorithms such as reinforcement learning and rejection sampling~\cite{liu2024large, wei2025swerl, yao2022react}. In summary, we make the following contributions:

\begin{enumerate}[leftmargin=*]
    \item We introduce a \emph{repository-level mutation framework} that can synthesize bug datasets at scale. The framework is designed to be extensible to other programming languages and scalable, generating large volumes of synthetic bugs with minimal manual intervention~\cite{ma2024lingmaswegptopendevelopmentprocesscentric}.
    \item We release \tool{}---our synthetic dataset comprising bug–fix pairs, test cases, and sampled repair trajectories.
    \item We empirically demonstrate that training LLM-based APR systems on synthetic data follows the same data scaling principles observed with real-world datasets~\cite{pan2024trainingsoftwareengineeringagents}. In our experiments, an LLM agent trained on \tool{} achieves a 15.3\% resolve rate on SWE-Bench Lite, compared to 13\% for systems trained on manually curated datasets. 
\end{enumerate}

Our results suggest that by leveraging realistic, process-aware synthetic data, LLM-based APR models can achieve state-of-the-art performance without the high costs associated with manual curation. Moreover, \tool offers a scalable solution to the lack of executable environments, one of the core bottlenecks in advancing open-source SWE agents. We hope this work contributes to democratizing progress in autonomous software engineering by making high-quality training data broadly accessible.
\newcommand{\mycode}{\lstinline}
\begin{figure*}[t]
    \centering
    \includegraphics[width=0.78\linewidth]{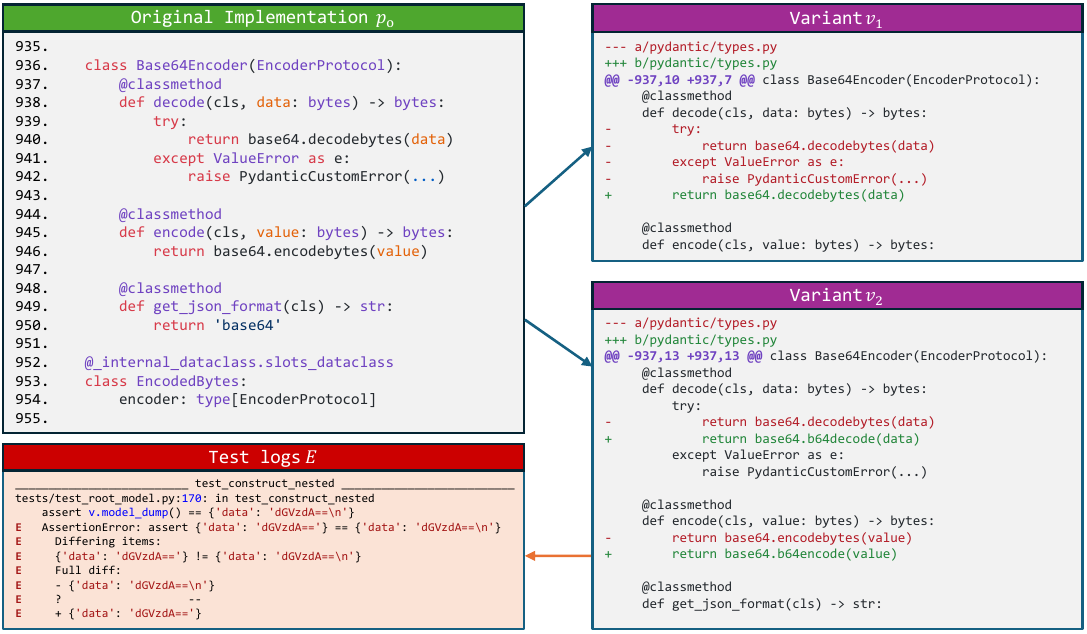}
    \vspace{-9pt}
    \caption{Illustrating Example of LLM-Generated Code}
    \label{fig:perfect-illustrating-example}
\end{figure*}

\section{Illustrating Example}
\label{sec:illustration}

This section illustrates how LLMs could be used to generate buggy code. 
In Figure~\ref{fig:perfect-illustrating-example}, we present an example on the \code{pydantic} library. The original code defines the \code{Base64Encoder} class, which contains three class methods: \code{decode}, \code{encode}, and \code{get\_json\_format}~\cite{pydantic}. We begin by masking the implementation details of these methods and then use LLM to regenerate them (Section \ref{sec:synbugs-data} for details). The resulting implementation often deviates from the original code, producing different variants, e.g., $v_1$ and $v_2$~(Figure~\ref{fig:perfect-illustrating-example}).

In the first variant \( v_1 \), the LLM-generated implementation omits the \code{try/except} block when calling \code{base64.decodebytes(data)}, leading to improper error handling when invalid data is passed. This oversight {\bf mirrors common human mistakes}, such as {\em forgetting to incorporate error-handling mechanisms before invoking specific APIs}. Such mistakes often stem from incomplete documentation or insufficient familiarity with the codebase requirements. 
This example also highlights that LLM-generated implementations can introduce bugs requiring \textbf{multi-line fixes}, such as adding a \code{try/except} statement and handling errors properly by raising an exception.

In the second variant \( v_2 \), the LLM's implementation replaces \code{encodebytes} and \code{decodebytes} with \code{b64encode} and \code{b64decode} from the \code{base64} library, deviating from the original code. This change causes the test \code{test\_construct\_nested} to fail due to subtle output differences between \code{b64encode} and \code{encodebytes}~\cite{so38214765}. This error exemplifies a {\bf typical API misuse}, where an incorrect API is selected \cite{amann2019systematic}, often due to a lack of domain knowledge about the \code{base64} library. Moreover, this LLM-generated error demonstrates the potential for introducing non-trivial bugs that require fixes across {\bf multiple locations}—reflecting the complexity of real-world debugs. Traditional mutation-based bug synthesis struggles to generate such multi-location bugs due to the combinatorial explosion of mutations, making LLM-based solutions a promising alternative.

These examples illustrate that, even within a single file, multiple {\em test-based verifiable} bugs can be generated synthetically. Moreover, different levels of granularity can be chosen for re-implementation, demonstrating the flexibility/scalability of LLM-based bug synthesis. Importantly, the examples emphasize that LLMs can effectively simulate {\em human errors, producing non-trivial, multi-hunk bugs}.

\begin{comment}
\footnote{\url{https://stackoverflow.com/questions/38214765/base64-encodebytes-v-b64encode-v-encodestring}}
\footnote{\url{https://github.com/pydantic/pydantic/blob/0a3135231d57597f47d48a9bb2b4b2ec9c46d63b/pydantic/types.py\#L936-L950}}
\end{comment}

\section{Problem Formulation}
\label{sec:formulation}

Let \( v \) denote a buggy program and let \( T = \{ t_1, t_2, \dots, t_n \} \) be its test suite. When \( v \) is executed on \( T \), it produces an error log \( E = \{ e_1, e_2, \dots, e_m \} \) capturing the failures. The goal of Automated Program Repair (APR) is to automatically generate a patch \( \Delta \) that, when applied to \( v \), produces a repaired program $p'$:
\[
p' = v \oplus \Delta
\]
that satisfies the specification encoded by \( T \):
\[
\text{Pass}(p', T) = \text{True}, i.e., \forall i=1..n, \text{Pass}(p', t_i) = \text{True}.
\]
Here, \( \oplus \) denotes the application of the patch \( \Delta \) to the buggy program \( v \). In a probabilistic framework, APR can be viewed as learning a mapping \( f \) parameterized by \( \theta \) that takes as input the buggy program \( v \) along with its error log \( E \) and outputs a patch \( \Delta \). That is, we aim to model APR via the conditional distribution
\[
P(\Delta \mid v, E)
\]
such that the generated patch \( \Delta \) repairs the program \( v \) (i.e., \( \text{Pass}(v \oplus \Delta, T) = \text{True} \)). The training data for APR is composed of tuples
\begin{equation}
\label{eq:delta}
    \mathcal{D} = \{ (v, T, E, \Delta) \},
\end{equation}
where each \( \Delta \) is a patch that transforms the buggy program \( v \) into one that passes all tests in \( T \). Let $\theta$ represent a sequence of modifications leading to the final patch \( \Delta \). In such cases, the learning objective can be formulated as maximizing the likelihood
\begin{equation}
\max_\theta \prod_{(v, T, E, \Delta, R) \in \mathcal{D}} P(\Delta \mid v, E; \theta).    
\end{equation}

Many APR approaches, however, focus solely on directly generating \( \Delta \) without explicitly modeling {\bf intermediate repair trajectory}. Such trajectories provide a step-by-step reasoning process, allowing the model to learn how developers iteratively fix bugs rather than just memorizing input-output mappings. Moreover, using such intermediate repair steps allows an APR model to learn structured, incremental fixes, leading to better generalization and improved interpretability. This is especially beneficial for learning complex bug fixes that require multiple interdependent changes. Training with such data makes it more aligned with how real-world debugging works where developers typically do not fix complex bugs in a single step but instead follow a process of incremental reasoning.

\section{LLM-based Variant Synthesis}
\label{sec:synbugs-data}

\begin{algorithm}[t]
\caption{Synthesize test-case verifiable variants with LLMs}
\label{algo:overview}
\begin{flushleft}
        \textbf{INPUT:} Original and correct program $p_o$, set of components $\mathcal{C}$ in program $p_o$, large language model $LLM$ \\
        \textbf{OUTPUT:} dataset of test-case verifiable variants $V$ \\
        \textbf{PROCEDURE:} 
\end{flushleft}
\begin{algorithmic}[1]
\STATE $V \gets $ [ ]
\WHILE{in budget}
    \STATE Sampling a component $c$ from program $p_o$
    \STATE $p_{\text{masked}}  \gets \text{maskComponent}(p_o, c)$
    \STATE Obtain the mutated component $\hat{c} \sim LLM ( \cdot \mid p_{\text{masked}})$
    \STATE $v \gets \text{replaceComponentInRepo}(p_{\text{masked}}, c, \hat{c})$

    \STATE T $\gets$ gatherTest($v$, $\hat{c}$)

    \IF {failed($v$, $T$)}
        \STATE Append $v$ to $V$
    \ENDIF
\ENDWHILE
\STATE return $V$
\end{algorithmic}
\end{algorithm}

Given a program \( p_o \) and its test cases, our goal is to generate multiple program variants, each of which fails at least one test case, thereby synthesizing bugs that resemble those made by humans. The process begins by selecting a component \( c \) from the original program \( p_o \). We then mask out its implementation, producing a modified version \( p_{\text{masked}} \). Next, we utilize LLMs to re-implement the masked component, generating a new version \( \hat{c} \). This newly synthesized component is then inserted back into \( p_{\text{masked}} \), forming a program variant \( v \). Finally, we retain only the variants that fail at least one test case.
The process of our synthesizing variants that can be verified by test cases is outlined in Algorithm~\ref{algo:overview}. The process consists of three main phases:
(1) component selection (lines 3--4), (2) leveraging LLM to re-implement the component (line 5), and (3) filtering variants by test cases (lines 8--10). 

%\vspace{-16pt}
\subsection{Component Selection} 
\label{subsection:component-selection}

In the component selection phase, our objective is to identify a component $c \in \mathcal{C}$ from the original program $p_{o}$ as the target for re-implementation. A component $c$ may be any program entity—such as a variable, statement, function, class, file, or module. This choice is pivotal: it determines where the synthetic bug will be injected and strongly influences the quality of the resulting variants. Selecting peripheral or rarely executed components often leads to trivial bugs that do not meaningfully test the program, whereas targeting central components, especially those covered by tests, produces variants that resemble realistic and impactful defects. We examine two strategies:  

{\em Naive Uniform Selection}: The simplest approach samples $c$ uniformly at random from $\mathcal{C}$, assigning $P(c) = \frac{1}{|\mathcal{C}|}$. This method is computationally cheap but assumes all components contribute equally to program behavior. In practice, programs are heterogeneous: some components (e.g., core utility functions) are critical, while others (e.g., corner-case statements) are rarely relevant. As a result, this strategy often yields less challenging bugs of limited practical value.  

{\em Test Coverage-Based Selection}: To address this limitation, we prioritize components according to their test coverage. Components executed by more test cases are more central to program behavior and more likely to contain bugs historically, making them better candidates for rewriting. This strategy increases the chance of generating synthetic bugs that are both impactful and representative of real-world defects. A practical challenge is the overhead of collecting coverage data, which requires running the test suite. To mitigate this, we cache coverage information for each repository commit and reuse it across multiple mutations, amortizing the cost and enabling efficient large-scale bug generation.

Formally, we model the relationship between test cases and components using a bipartite graph:
$G = (T \cup C, E),$ where
\begin{itemize}
    \item \( T \) represents the set of test cases associated with \( p_o \),
    \item \( C \) denotes the set of components in \( p_o \),
    \item \( E \) is the set of edges, with each edge  \( (t, c) \in E \) indicate that test case \( t \in T\) exercises/covers the component \(c \in C\).

\end{itemize}

For a component \( c \in C \), its degree in the graph $G$, denoted by \(\deg(c)\), is the number of test cases that cover it: $\deg(c) = \left|\{ t \in T \mid (t, c) \in E \}\right|$.
We interpret \(\deg(c)\) as a quantitative measure of the component's importance. A high degree suggests that \( c \) is integral to multiple aspects of the program's behavior, as it is scrutinized by more test cases. Conversely, a low degree indicates a component with limited influence, deliberately confined to a specific use case, or rarely executed path. Using this metric, we define the probability of selecting \( c \) as proportional to its test coverage:
\[
P(c) = \frac{\deg(c)}{\sum_{c_i \in C} \deg(c_i)}.
\]
Here, the denominator normalizes the probabilities across all components, ensuring that
$\sum_{c \in C} P(c) = 1.$ The component \( c \) is then sampled from this derived  distribution: $ c \sim P$. This weighted sampling biases the selection toward components with higher test coverage, aligning the component selection with the program's critical regions with creating more challenging bugs, therefore, valuable for training. We validate this hypothesis in Table~\ref{tab:component-selection} in Section~\ref{sec:component-selection-exp}.

\subsection{LLM-based Re-implementations as Variants}
\label{sec:llm-reimplementation}

\begin{figure}[t]
\small
\begin{tcolorbox}[colback=gray!5!white, colframe=gray!75!black]
    
Given the code below, please implement the body of the function \textcolor{orange}{\{function\_name\}} directly without changing surrounding context. \\
\textcolor{orange}{\{file\_content\_with\_function\_body\_removed\}} 
\\
Answer in the following format:\\  
<explain your implementation>  
\begin{verbatim}
```
\end{verbatim}
\textcolor{orange}{\{function\_signature\}}\\
\hspace*{2em}... your code goes here ...
\begin{verbatim}
```
\end{verbatim}
\end{tcolorbox} 
\vspace{-12pt}
    \caption{LLM Prompt to re-implement a function}
    \label{fig:reimplementation-prompt}
\end{figure}

In this work, we leverage the LLMs' ability to understand both the surrounding context and the component specification to generate the re-implementation. For each selected component—whether a function, method, or class—we first remove its internal implementation (i.e., masking it) while preserving its interface and documentation. We then prompt the LLM with a structured query to synthesize a variant of the component. Figure~\ref{fig:reimplementation-prompt} illustrates a typical prompt, which includes the file content with the masked component, the component's signature, and its name, thereby guiding the LLM to generate a semantically coherent variant.

Let $c \in \mathcal{C}$ be a component selected from the set of all components. We corrupt the original component by masking its implementation details while retaining its signature and docstring. 
For brevity, in Algorithm \ref{algo:overview}, we simplify the notation for invoking LLM as:
\[
\hat{c} \sim LLM(\cdot \mid p_{\text{masked}}),
\]
where $p_{\text{masked}}$ is a shorthand for the combined context $(p^c_{\text{masked}}, s, p_s)$
\begin{itemize}
    \item $p^c_{\text{masked}}$ denotes the contextual information surrounding target component $c$ and derived from $p_{\text{masked}}$ ,
    \item $s$ corresponds to specifications of original component $c$, and
    \item $p_s$ is the structured prompt provided to the LLM.
\end{itemize}
Because the LLM is not conditioned on the original implementation $p_o$, the generated variant $\hat{c}$ may differ significantly—introducing subtle, realistic modifications that emulate human-like bugs. We have different reimplementation strategies for components:

{\em For a function, method:} we remove its original body, leaving the signature and docstring intact, and prompt the LLM to generate a new implementation that adheres to the given context. For example, as re-implementing a function that filters a list based on specific criteria, the LLM might modify the filtering logic (e.g., reordering conditions or altering comparison operators) in a way that still follows the documented intent yet introduces a non-trivial bug.

{\em For a class:}  we clear the bodies of all methods while preserving the class attributes and documentation, providing the entire file context to the LLM. For instance, in re-implementing a class that manages database operations, the LLM might adjust the sequence of method calls or modify error handling mechanisms. The resulting variant remains structurally consistent with the original interface but may incorporate subtle logical flaws that affect overall behavior.

Finally, after re-implementation, each variant of a component is integrated back into the original codebase to form a complete variant. We then execute the associated test cases and retain only those variants that fail at least one test case. This execution-based filtering is essential for discarding non-buggy variants.

\section{Ground-truth Extraction for Supervised Fine-Tuning}
\label{sec:groundtruth-extraction}

In this section, we explain our ground-truth extraction process.

Given a set of variants $V_i$ generated for each program $p_i$ using Algorithm~\ref{algo:overview}, our objective is to construct a training dataset according to Equation~(\ref{eq:delta}):
\begin{equation}
    \mathcal{D} = \{ (v_{ij}, T_i, E_{ij}, \Delta_{ij}) \mid p_i \in \mathcal{P},\, v_{ij} \in V_i \},
\end{equation}

where:
\begin{itemize}
    \item $\mathcal{P}$ is the set of well-tested, correct programs (repositories),
    \item $V_i$ denotes the set of variants generated for the program $p_i$,
    \item $v_{ij}$ denotes a buggy variant generated via our re-implementa\-tion process (with $j$ indexing the variants),
    \item $T_i$ is the set of test cases for program $p_i$,
    \item $E_{ij}$ is the error log produced when $v_{ij}$ is executed against $T_i$, and
    \item $\Delta_{ij}$ is the patch that, when applied to $v_{ij}$, converts it into a version that passes all test cases.
\end{itemize}

In the supervised fine-tuning (SFT) procedure, {\em the model is provided with a buggy variant $v_{ij}$ and its corresponding error log $E_{ij}$, and it is trained to generate the patch $\Delta_{ij}$ that repairs the bug such that the modified program passes all test cases}. Optionally, the dataset can also include {\em intermediate repair steps $R$, representing the sequence of modifications} made during the debugging process. This addition aims to emulate the iterative, step-by-step approach that developers typically follow when resolving issues.

\paragraph{Naive Approach: Reverse Patch Diff}  
A straightforward method to obtain $\Delta_{ij}$ is by computing the reverse patch diff between the original program $p_i$ (which passes all tests) and the variant $v_{ij}$:

\[
\Delta_{ij} = \text{diff} (p_i , v_{ij}).
\]
While this approach provides a direct ground-truth patch for training, it does not capture the {\em iterative repair process} and thus offers limited incremental context. Furthermore, we also found that sometimes, our rewriting process produces far different implementation compared with the original, leading $\Delta_{ij}$ to be an un-natural fix. 
In Table~\ref{tab:reverse-patch-diff-performance} (Section~\ref{sec:rq8}) we show that training on this naive representation alone does not yield optimal results and provides low-quality training patches. One could consider implementing guided intermediate steps toward the reverse patch diff by providing hints during repairing process \cite{zelikman2022star}; however, such a guided rollout risks leaking information about the ground-truth (e.g location of the bug) and is difficult to achieve final patch exactly equals to $\Delta_{ij}$ due to stochastic essence of sampling LLM's intermediate steps, resulting in only a small number of usable samples. Moreover, devising a procedure that avoids such leakage remains unclear, and we leave these investigations for future work.

\paragraph{Training on Success Rollout Trajectories}  
Alternatively, we derive $\Delta_{ij}$ from a successful rollout trajectory of repair actions. In this strategy, an LLM agent operates on $v_{ij}$ and performs a series of incremental steps of codebase exploration or modification:
\[
R_{ij} = \{r^{ij}_1, r^{ij}_2, \dots, r^{ij}_K\},
\]

Each $r_k$ can produce the corresponding observation $o_k$, e.g., the function body if $r_k$ is a code search action, however, for simplicity, we omit observations from the maximum likelihood formula and our notation, details can be referred to \cite{xi2024agentgym}.

These intermediate steps culminate in a final patch $\Delta_{ij}$ that, when applied, produces a program version that passes all test cases. The agent is embedded in an environment that mirrors a real-world debugging workflow, allowing actions such as code modifications or codebase searches at each step. Only those trajectories where the final patch is verified by the test suite are retained, therfore resulting in an augmented version of $\mathcal{D}$ in Equation (3), $\mathcal{D_{\text{aug}}}$:
\begin{equation}
\mathcal{D}_{\text{aug}} = \{ (v_{ij}, T_i, E_{ij}, \Delta_{ij}, R_{ij}) \mid p_i \in \mathcal{P},\, v_{ij} \in V_i \},    
\end{equation}
We fine-tune the LLM parametrized by $\theta$ by maximizing: 
\begin{equation}
    \max_\theta \prod_{(v_{ij}, T_i, E_{ij}, \Delta_{ij}, R_{ij}) \in \mathcal{D}_{\text{aug}}} P(\Delta_{ij}, R_{ij} \mid v_{ij}, E_{ij}; \theta).
\end{equation}

which is a stricter version of Equation (2) but enriches the supervisory signal by providing a sequence of context-aware repair steps. This also better simulates how developers iteratively debug and refine their code.

\section{{\tool}: A Synthetic Dataset Reflecting Real-World Bug-Fix Patterns}
\label{sec:dataset-synbug}
\begin{table}[ht]
    \centering
    \tabcolsep 1.5pt
    \small
    \caption{{\tool} Statistics. \# of variants and \# of patches are counted in total. Number of fail tests and the test log's length (measured in tokens) are reported by median across variants. Other columns are the average value across variants. Number of steps per trajectory are \# actions in MoatlessTool.}
    \label{tab:data-statistics}
    \vspace{-6pt}
\begin{tabular}{@{}c|rrrrrrrr@{}}
\toprule
Repo & \multicolumn{1}{c}{\begin{tabular}[c]{@{}c@{}}Total \#\\ variants\end{tabular}} & \multicolumn{1}{c}{\begin{tabular}[c]{@{}c@{}}\# Fix\\ patches\end{tabular}} & \multicolumn{1}{c}{\begin{tabular}[c]{@{}c@{}}\# Lines\\ edited\end{tabular}} & \multicolumn{1}{c}{\begin{tabular}[c]{@{}c@{}}\# Diff \\ hunk\end{tabular}} & \multicolumn{1}{c}{\begin{tabular}[c]{@{}c@{}}\# Files\\ edited\end{tabular}} & \multicolumn{1}{c}{\begin{tabular}[c]{@{}c@{}}\# Fail\\ tests\end{tabular}} & \multicolumn{1}{c}{\begin{tabular}[c]{@{}c@{}}Test log\\ length\end{tabular}} & \multicolumn{1}{c}{\begin{tabular}[c]{@{}c@{}}\# Steps\\ per traj\end{tabular}} \\ \midrule
bokeh & 623 & 449 & 4.95 & 1.22 & 1.08 & 3 & 4414 & 9.34 \\
conan & 1525 & 383 & 6.00 & 1.36 & 1.05 & 8 & 9361 & 8.32 \\
dask & 691 & 103 & 7.87 & 1.22 & 1.02 & 18 & 67398 & 7.43 \\
hydra & 2525 & 745 & 5.11 & 1.18 & 1.04 & 13 & 93632 & 8.65 \\
moto & 222 & 93 & 5.68 & 1.31 & 1.04 & 3 & 14108 & 9.28 \\
dvc & 1607 & 645 & 6.34 & 1.25 & 1.06 & 6 & 42632 & 8.62 \\
pydantic & 2266 & 600 & 5.88 & 1.27 & 1.09 & 9 & 3802 & 8.64 \\ \midrule
Summary & 9459 & 3018 & 5.73 & 1.25 & 1.06 & 8 & 18842 & 8.68 \\ \bottomrule
\end{tabular}
\end{table}

This section presents {\tool}, our high-quality synthetic dataset constructed using the synthesis methodology described in our previous section.

\subsubsection*{\bf Procedure and Models}
We construct the repository set \(\mathcal{P}\) for {\tool} based on SWE-Gym~\cite{pan2024trainingsoftwareengineeringagents}, which includes 11 Python repositories containing 2,438 tasks and well-maintained test suites. Each task involves submitting a patch to fix a real bug. While SWE-Gym is Python-specific, our re-implementation and ground-truth extraction pipeline is language-agnostic and generalizable to other codebases. Due to computational constraints, we selected seven repositories for our experiments (see Table~\ref{tab:data-statistics}).

From each selected repository, we sampled five committed program versions as original seeds \( p_o \). Using our mutation methodology (Algorithm~\ref{algo:overview}), we generated program variants by injecting mutations at the function and class level. For component selection, we applied two strategies: (i) uniform random sampling and (ii) test coverage-based selection, where components were mapped to test cases using line coverage metrics obtained via \code{coveragepy}~\cite{coveragepy}.

To generate variants, we employed Qwen 2.5 Coder Instruct 32B~\cite{hui2024qwen25codertechnicalreport}, using a temperature of 0.7 and the prompt template shown in Figure~\ref{fig:reimplementation-prompt}. We discarded variants that either compiled but passed all test cases or failed to compile. All experiments were executed in a containerized environment on a machine with 48 CPU cores over a 120-hour period.

\subsubsection*{\bf Synthesizing Ground-Truth Fixes for Supervised Fine-Tuning}
\label{sec:synbugs-sft-statistic}

To enable supervised fine-tuning (SFT), we extracted ground-truth patches from successful rollout trajectories (Section~\ref{sec:groundtruth-extraction}). While our framework is compatible with various agent scaffolds, we used the open-source MoatlessTool~\cite{moatless_tool}, which integrates code search with LLM-based retrieval. Each variant’s failed test log was truncated to 16k tokens and included in the input prompt. For embedding-based retrieval, we used Jina Embeddings v2 Base Code~\cite{günther2024jinaembeddings28192token}.

We sampled repair trajectories using Qwen 2.5 Coder Instruct 32B in three rounds: one deterministic pass (temperature = 0) and two stochastic passes (temperature = 1). The generated patch from each rollout was applied to the variant, and successful patches—i.e., those that passed all associated test cases—were accepted. These valid rollouts were added to {\tool}, and the corresponding patches were recorded as ground-truth fixes.

\subsubsection*{\bf Dataset Statistics}
\label{sec:synthesized-dataset-statistic}

Table~\ref{tab:data-statistics} summarizes the key statistics of {\tool}. We generated 21,804 program variants from 35 original program versions across seven repositories. After filtering invalid or unverifiable variants, {\tool} contains 9,459 high-quality, test-case-verifiable program variants.

\begin{figure}[t]
    \centering
    \includegraphics[width=5.5in]{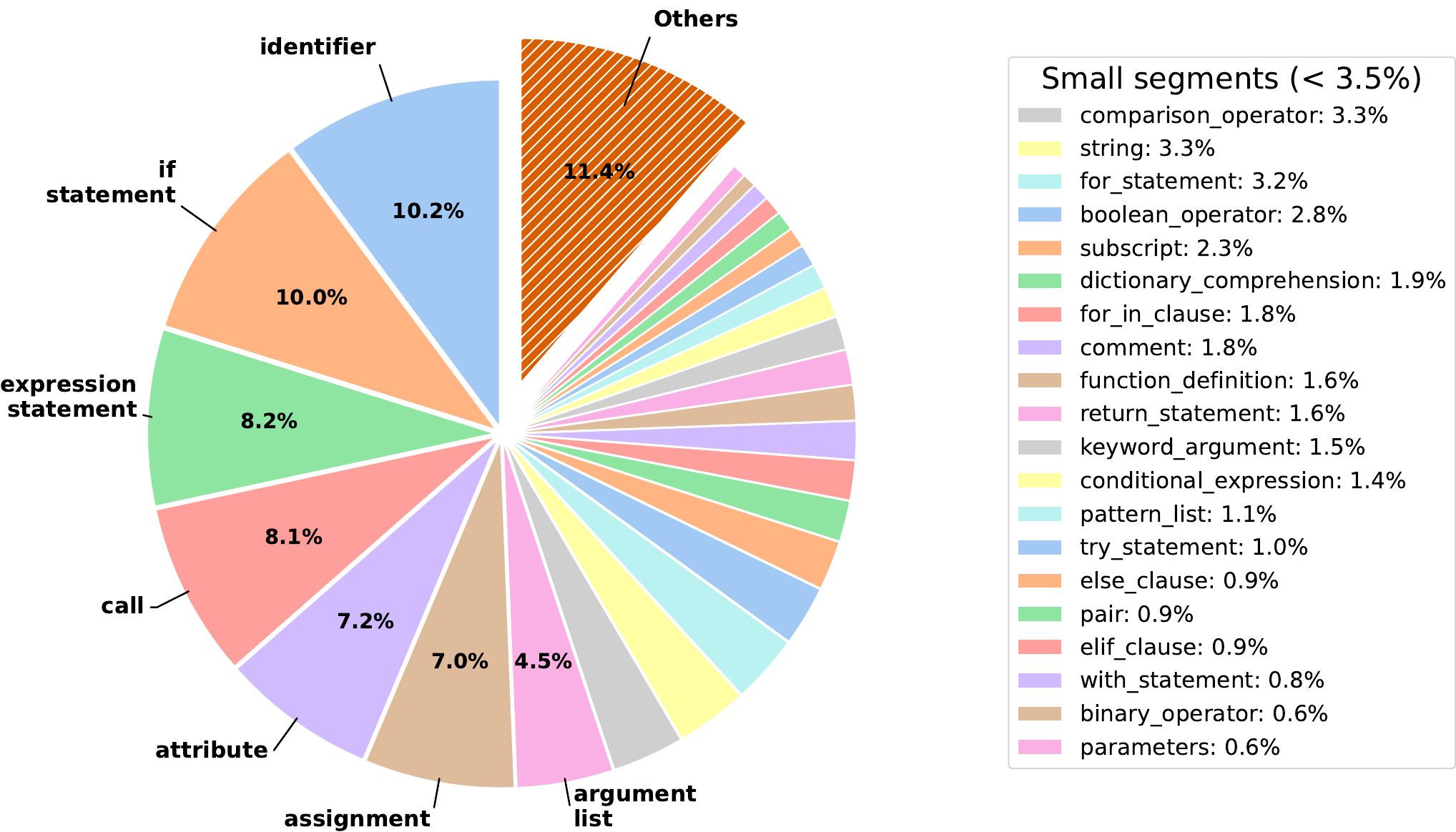}
    \caption{Diversity of Fixing Locations in {\tool}}
    \label{fig:fixed-stmts}
\end{figure}

Two component selection strategies were used: test coverage-based selection (40.88\%, 3,867 variants) and uniform random sampling (59.12\%, 5,592 variants). Of the 9,459 variants, 7,798 remain buggy, while 1,661 led to successful fixes through trajectory rollouts, yielding 3,018 patched program versions. The overall fix success rate is 10.6\%. Columns 4–6 of Table~\ref{tab:data-statistics} provide patch-level insights:

\begin{itemize}[leftmargin=*]
    \item \textbf{Patch complexity.} Each patch modifies an average of 5.73 lines, spread across 1.25 hunks and 1.06 files, indicating that bug fixes are rarely confined to single-line edits. This reflects the complexity of real-world software maintenance, where patches often involve multi-location changes.

    \item \textbf{Fix diversity.} Figure~\ref{fig:fixed-stmts} shows that bugs span diverse code locations, reinforcing the generalizability of {\tool} across various bug types and contexts.

    \item \textbf{Test signal richness.} Each buggy variant is associated with 254.9 test cases on average (median: 72), including a median of 8 failing test cases. This provides a strong and realistic supervision signal for LLM training.

    \item \textbf{Failure log length.} The median length of test failure logs is 18,842 tokens, capturing detailed execution traces that enable APR models to learn from rich contextual information—mimicking how developers analyze logs to fix bugs.

    \item \textbf{Repair trajectory richness.} Each fix includes an average of 8.68 intermediate repair steps. These fine-grained trajectories allow models to learn step-wise debugging strategies, improving planning and robustness on unseen bug scenarios.
\end{itemize}

\section{Empirical evaluation}

\subsection{Research Questions}
For evaluation, we seek to answer the following research questions:

\noindent {\em (1) Manual data vs synthetic data}:

\textbf{$RQ_1$. Model performance comparison on synthetic and manual data: } How do the manual and synthetic training data in {\tool} influence the performance of models, when the training data is controlled to have either (a) the same total number of variants, or (b) the same total number of trajectories? 

\textbf{$RQ_2$. Synthetic Data Scaling:}  How does increasing the number of synthetic training instances affect model performance?

\textbf{$RQ_3$. Human Study:}  How well can human subjects distinguish {\tool}'s results from real-world, manually collected bugs?

\noindent {\em (2) Can {\tool}'s synthetic data improve models across size?} 

\textbf{$RQ_4$. Model Performance with Different Model Sizes:} How does model performance vary across different model sizes when being fine-tuned on our synthetic training data in {\tool}?

\noindent {\em (3) In-depth and ablation study of our pipeline:}

\textbf{$RQ_5$. Impact of component granularity:} How do different component granularities affect the trained models' performance?

\textbf{$RQ_6$. Impact of component selection:} How do different component selection strategies affect the trained models’ performance?

\textbf{$RQ_7$. Impact of model size:} How does the size of the model used for component rewriting affect the trained models' performance?

\textbf{$RQ_8$. Ground-truth extraction strategies:} How well does the model perform when being trained on reverse patch diff compared to that when being trained on {\tool} with rollout?

\subsection{Empirical Methodology}

\subsubsection{Models, Benchmarks, and Procedure}

We chose the Qwen 2.5 Coder Instruct models~\cite{hui2024qwen25codertechnicalreport} with different sizes 7B, 14B, and 32B. After training the models with {\tool} and the manual dataset in \cite{pan2024trainingsoftwareengineeringagents}, we evaluated them primarily on
the SWE-Bench Lite \cite{jimenez2024swebench} dataset, which includes 300 bugs from 12 popular Python repositories, each having a patch based on its bug report.
We also evaluate on the SWE-Bench Verified \cite{chowdhury2024swebenchverified} dataset, which includes 500 bugs from the same set of repositories, but with human validation to ensure that each issue contains sufficient information to be solvable. Additionally, we assess performance on the BugsInPy~\cite{BugsInPy} dataset, which contains 493 bugs from 17 projects, though we successfully reproduced only 317 bugs (due to missing required libraries).
To align with our problem formulation for APR (see Section~\ref{sec:formulation}), we use failed test logs from these datasets as task prompts.

All training experiments were conducted using LoRA \cite{hu2021loralowrankadaptationlarge} with a rank of 8, an alpha of 16, and a learning rate of $2\times10^{-5}$. Training was performed on 4 H100 GPUs for 5 epochs with a global batch size of 8. We utilized the LLAMA-Factory framework \cite{zheng2024llamafactoryunifiedefficientfinetuning} and incorporated LongLoRA \cite{chen2023longlora} to support a context length of 65,536 during training. Unless otherwise specified, we fine-tuned from Qwen 2.5 Coder Instruct \cite{hui2024qwen25codertechnicalreport} with a model size of 14B.
% For the 32B model, we employed QLoRA \cite{dettmers2023qloraefficientfinetuningquantized} for fine-tuning.

\subsubsection{Evaluation Metrics}

We evaluate the automated program repair (APR) models trained on different datasets using three key metrics: {\em Resolve Rate}, {\em Correct Patch Rate}, and {\em Empty Patch Rate}. We counted a patch generated by an APR model toward resolve rate if it passes all test cases. It is correct if its abstract syntax tree (AST) matches the reference developer patch in the ground truth (ignoring formatting differences). The empty patch rate measures the percentage of predictions that present no fixing, a phenomenon usually occurring with LLM-based agents. A higher resolve rate and correct patch rate indicate better performance of an APR model trained on a dataset, whereas a lower empty patch rate is desirable.

\subsection{Manual Data versus Synthetic Data }

\begin{table}[t]
    \centering
    \small
    \tabcolsep 2.5pt
    \caption{Performance when training on 230 data instances from SWE-Gym (manual) compared to that training on {\tool} (synth), with different per-bug instance caps (RQ1)}
    \label{tab:cap-table}
    \vspace{-6pt}
\begin{tabular}{@{}cc|rrrrrr@{}}
\toprule
\multicolumn{2}{c|}{\multirow{2}{*}{Dataset}} & \multicolumn{2}{c}{$\%$ Resolved $(\uparrow)$} & \multicolumn{2}{c}{$\%$ Correct $(\uparrow)$} & \multicolumn{2}{c}{$\%$ EP $(\downarrow)$} \\ \cmidrule(l){3-8} 
\multicolumn{2}{c|}{} & \multicolumn{1}{c}{manual} & \multicolumn{1}{c}{synth} & \multicolumn{1}{c}{manual} & \multicolumn{1}{c}{synth} & \multicolumn{1}{c}{manual} & \multicolumn{1}{c}{synth} \\ \midrule
\multicolumn{1}{c|}{\multirow{3}{*}{\begin{tabular}[c]{@{}c@{}}SWE-Bench\\ Lite\end{tabular}}} & Cap 1 & 12.00\% & 9.67\% & 1.33\% & 1.67\% & 29.33\% & 38.33\% \\
\multicolumn{1}{c|}{} & Cap 2 & 11.67\% & 12.00\% & 1.67\% & 1.33\% & 28.33\% & 28.00\% \\
\multicolumn{1}{c|}{} & Cap 3 & 12.00\% & 11.67\% & 1.67\% & 1.00\% & 28.00\% & 31.67\% \\ \midrule
\multicolumn{1}{c|}{\multirow{3}{*}{BugsInPy}} & Cap 1 & 16.09\% & 19.24\% & 1.58\% & 1.89\% & 29.65\% & 27.13\% \\
\multicolumn{1}{c|}{} & Cap 2 & 16.72\% & 17.67\% & 1.58\% & 1.89\% & 33.75\% & 28.39\% \\
\multicolumn{1}{c|}{} & Cap 3 & 17.03\% & 18.61\% & 2.21\% & 2.21\% & 28.39\% & 27.44\% \\ \bottomrule
\end{tabular}
\end{table}

%\vspace{-15pt}
\subsubsection{Model performance comparison on synthetic and manual data}

\textbf{$RQ_1$: How do models perform when being trained on SWE-Gym manual dataset and {\tool}?}
Our goal is to compare the models' performance when being trained on manually curated bug-fix data (sourced from GitHub such as SWE-Gym Lite) versus synthetic bug-fix data in {\tool} on SWE-Bench Lite and BugsInPy benchmarks. 
A bug-fix dataset affects APR performance via three interrelated factors: (1) {\em Diversity:} the number of unique bugs (variants) available; (2) {\em Dataset size:} the total number of success trajectories obtained during rollouts; and (3) {\em Per-Bug instance capping:} the maximum number of success trajectories per bug. Per-bug instance capping is especially critical because the repeated sampling in the rollout phase tends to favor easier bugs—since these yield more successful trajectories—while potentially under-representing harder bugs. {\em A very low cap might restrict the training data too severely, whereas a high cap could skew the training distribution towards easier tasks}. To tease these effects apart, we established two settings in this experiment:

\begin{table}[t]
\centering
\small
\tabcolsep 1.6pt
\caption{APR Performance of Qwen 2.5 Coder Instruct 14B fine-tuned with 1K success trajectories from 32B (RQ1)}
\label{tab:same-num-traj}
\vspace{-6pt}

\begin{tabular}{@{}ccccccc@{}}
\toprule
\multirow{2}{*}{Dataset} & \multicolumn{2}{c}{$\%$ Resolved $(\uparrow)$} & \multicolumn{2}{c}{$\%$ Correct $(\uparrow)$} & \multicolumn{2}{c}{$\%$ EP $(\downarrow)$} \\ \cmidrule(l){2-7} 
 & manual & synth & manual & synth & manual & synth \\ \midrule
SWE-Bench Lite & 11.33\% & 11.67\% & 2.33\% & 1.33\% & 33.00\% & 25.67\% \\ \midrule
SWE-Bench Verified & 11.40\% & 11.00\% & 1.60\% &	1.80\% &	31.40\% &	28.80\% \\ \midrule
BugsInPy & 19.56\% & 17.67\% & 1.89\% & 1.89\% & 28.08\% & 28.08\% \\ \bottomrule
\end{tabular}

\end{table}

\begin{table}[t]
\centering
\small
\tabcolsep 1.6pt
\caption{APR Performance of Qwen 2.5 Coder Instruct 32B fine-tuned with 2136 success trajectories from DeepSeek-v3-0324 (RQ1)}
\label{tab:same-num-traj-deepseek}
\vspace{-6pt}

\begin{tabular}{@{}ccccccc@{}}
\toprule
\multirow{2}{*}{Dataset} & \multicolumn{2}{c}{$\%$ Resolved $(\uparrow)$} & \multicolumn{2}{c}{$\%$ Correct $(\uparrow)$} & \multicolumn{2}{c}{$\%$ EP $(\downarrow)$} \\ \cmidrule(l){2-7} 
 & manual & synth & manual & synth & manual & synth \\ \midrule
SWE-Bench Lite & 23.67\% & 24.33\% & 3.33\% & 2.00\% & 13.00\% & 5.00\% \\ \midrule
SWE-Bench Verified & 18.20\% & 22.20\% & 3.80 \% & 3.20\% & 17.80\% & 8.00\% \\ \bottomrule
\end{tabular}
\end{table}

{\em Same number of variants}: We fixed the sample size to 230 instances from SWE-Gym Lite and an equal number from {\tool}, each undergoing three rollout rounds. Supervised fine-tuning was performed on success trajectories with varying caps on per-bug instances. As shown in Table \ref{tab:cap-table}, under the most lenient cap (Cap 3), \tool’s advantage is clear. While overall resolve rates on SWE-Bench Lite are similar (12.00\% vs. 11.67\%), \tool consistently outperforms on BugsInPy, with the largest gap at Cap 1 (19.24\% vs. 16.09\%). This suggests \tool provides a richer training signal through controlled diversity and balanced bug difficulty.

{\em Same total number of trajectories}: Here, we standardized training data to 1,000 success trajectories per dataset, without capping or variant constraints. Table~\ref{tab:same-num-traj} shows the two datasets converge: {\tool} matches or slightly surpasses manual data on SWE-Bench Lite (11.67\% vs. 11.33\%). On BugsInPy, both achieve the same correctness (1.89\%) and empty patch rates (28.08\%), with manual data slightly ahead in resolve rate (19.56\% vs. 17.67\%). Thus, while manual data benefits from real-world quality, the scalability and diversity of \tool compensate for this, yielding comparable performance.

To further test rollout effects under fixed trajectory budgets, we used \code{DeepSeek-v3-0324} \cite{deepseek-ai2024deepseek0v3} with $\code{CodeXEmbed}_{\text{2B}}$ \cite{liu2024codexembed0} to sample 2,136 success trajectories per dataset, fine-tuning \code{Qwen 2.5 Coder Instruct 32B} (context length 32,768). The baseline resolve rate on SWE-Bench Verified was 16.4\%. As shown in Table~\ref{tab:same-num-traj-deepseek}, training on \tool boosted this to 22.20\%, clearly surpassing manual data (18.20\%). This improvement stems from \tool’s larger set of bug instances and trajectory diversity, suggesting the synthetic advantage grows with larger fine-tuning data and stronger teacher models.

\subsubsection{Synthetic Data Scaling}
\label{sec:rq2}

\begin{figure}
    \centering
    \includegraphics[width=0.92\linewidth]{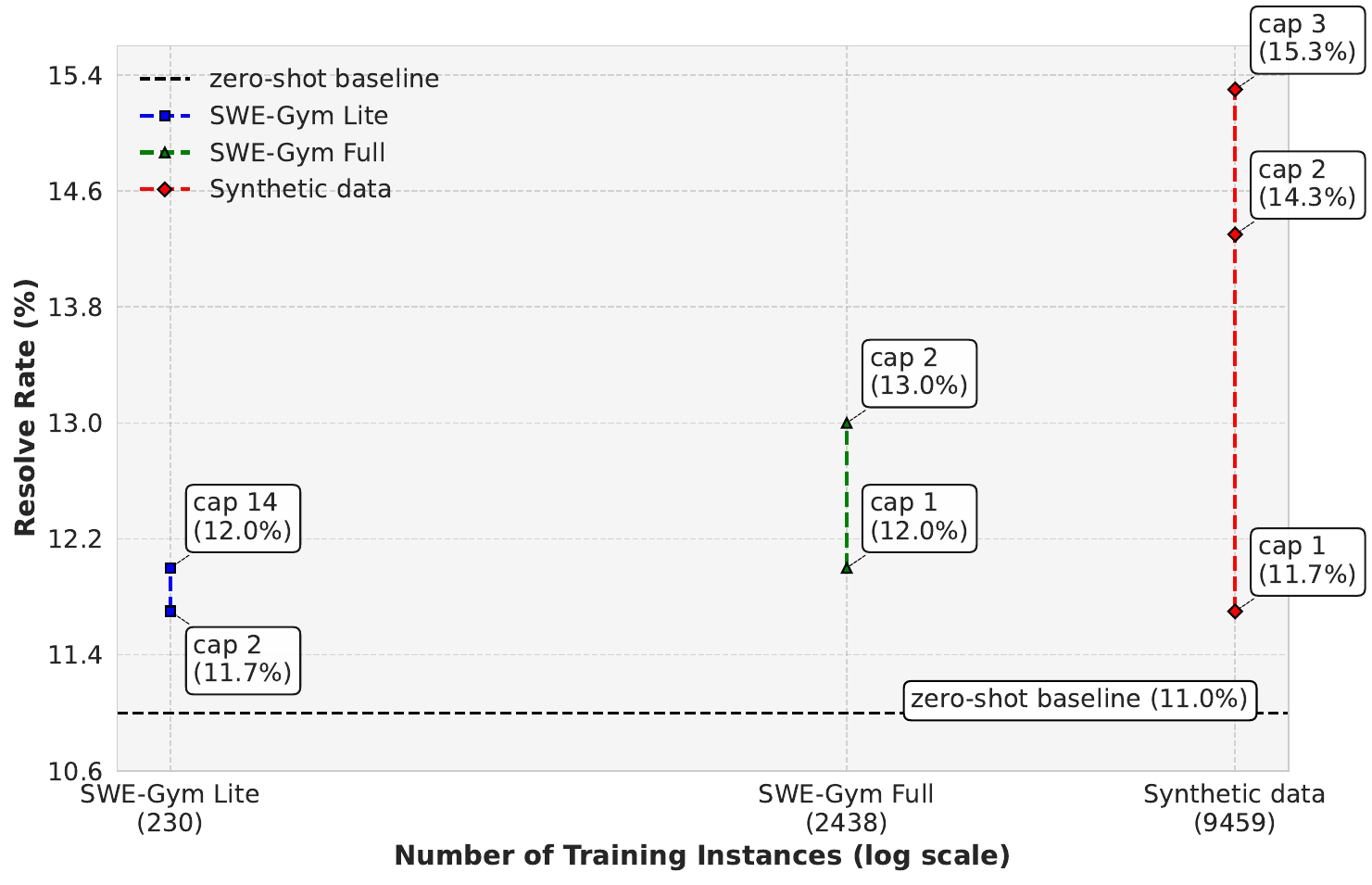}
    \vspace{-9pt}
    \caption{APR performance of models trained on synthetic vs real-world bugs evaluated on SWE-Bench Lite logs, finetuned in Moatless settings with Qwen2.5-Coder-14B Instruct (RQ2)}
        \label{fig:beyond-the-boundary}
\end{figure}

\textbf{$RQ_2$: How does increasing the number of synthetic training instances affect model performance?}
In this experiment, our primary motivation is to determine whether more—and more diverse—bug-fix examples can consistently yield improvements in an APR model’s ability to generate correct patches. To this end, we gradually increase the training set size from the size of SWE-Gym Lite (230 instances) to SWE-Gym Full's (2,438 instances) and finally to {\tool}'s (9,459 instances) to examine the impact of data scaling on APR performance without any constraint on number of unique bugs and number of trajectories.~As seen in Figure \ref{fig:beyond-the-boundary}, while the zero-shot baseline is around 11\%, fine-tuning on SWE-Gym Lite provides a modest boost (up to 12\%), and that on SWE-Gym Full further improves performance (up to 13\%). Notably, training with {\tool}, the model achieves the highest resolve rate (15.3\%)—surpassing all previous configurations. 

A noteworthy trend is that {\tool} exhibits a pronounced improvement when the cap on the number of per-bug instances is increased  (e.g., from Cap 1 to Cap 3) compared to smaller datasets. A plausible explanation is that with large caps, \tool’s larger and more diverse bug distribution reduces overfitting on simpler tasks, enabling APR model to effectively leverage additional success trajectories without overshadowing complex or less frequent bugs.

%===============================================
\begin{table}[t]
    \centering
    \small
    \caption{Human Study Statistics (RQ3): 11 participants}
    \label{tab:humanstudy-stats}
    \vspace{-6pt}
    \begin{tabular}{l r}
        \toprule
        \textbf{Metric} & \textbf{Value} \\
        \midrule
%        Total participants & 11 \\
        Total responses & 377 \\
        Overall accuracy & 55.17\% \\
        Synthetic accuracy & 45.11\% \\
        Real accuracy & 64.77\% \\
        \midrule
        True Positives (SYN $\rightarrow$ SYN) & 83 \\
        False Positives (REAL $\rightarrow$ SYN) & 68 \\
        False Negatives (SYN $\rightarrow$ REAL) & 101 \\
        True Negatives (REAL $\rightarrow$ REAL) & 125 \\
        \bottomrule
    \end{tabular}
    %\vspace{-9pt}
\end{table}

\subsubsection{Human Study} RQ\textsubscript{3}: \textbf{How well can humans distinguish \tool{}'s results from real-world bugs?}

\paragraph{Study Design.}
To evaluate the realism of our synthetic bugs, we conducted a \textit{web-based survey}~\cite{survey} with 11 participants experienced in Python. Each participant was presented with a total of 400 bugs—200 real bugs sampled from GitHub issues and 200 synthetic bugs generated by \tool{}. For each bug, we provided a buggy code snippet along with its test-case validated fix and corresponding error log trace. In addition, a brief description of the bug, generated by Claude 3.7 Sonnet \cite{tools}, was included to help participants quickly grasp the core issue and its context. Participants then labeled each bug as “Real” or “Synthetic”. In total, 377 responses were collected, with each participant evaluating 34 bugs on average.

\paragraph{Results.}
We computed accuracy (percentage of correct labels), synthetic accuracy, real accuracy, and confusion metrics. As seen in Table~\ref{tab:humanstudy-stats}, overall accuracy was 55.17\%, only slightly above random guessing, with real bugs correctly identified 64.77\% of the time and synthetic bugs just 45.11\%. Synthetic bugs were more frequently misclassified as real (101 false negatives) than real bugs misclassified as synthetic (68 false positives), showing that participants found our synthetic bugs difficult to distinguish from real ones.~The results in Table \ref{tab:humanstudy-stats} suggest that participants recognized genuine bugs reasonably well. {\tool} also synthesizes bugs realistic enough for effective APR. Readers are encouraged to take our survey~\cite{survey}.

%=============================================
\subsection{Synthetic Data and Model Size}

\subsubsection*{Impact of Model Size on {\tool}}

\textbf{$RQ_4$: How does model performance change across different model sizes when fine-tuned on our synthetic training data?}

We investigate whether smaller models gain disproportionately from additional bug-fix data or if larger models still see comparable improvements. We fine-tuned Qwen 2.5 Coder models with 7B, 14B, and 32B parameters on 1,027 success trajectories, using QLoRA \cite{dettmers2023qloraefficientfinetuningquantized} for the 32B model.

Table~\ref{tab:train-model-size} shows that the 7B model benefits most, with a sharp drop in empty patch rate and a clear rise in resolve rate over its weak zero-shot baseline. For larger models, gains narrow since the training trajectories—generated by Qwen 2.5 Coder Instruct 32B—already provide a strong signal, effectively distilling knowledge from strong teacher to smaller students. This Weak-to-Strong effect \cite{bansal2024smaller} diminishes with increasing student size, consistent with prior work \cite{pan2024trainingsoftwareengineeringagents, bansal2024smaller}. Addressing this may require self-improvement methods, such as Reinforcement Learning with Verifiable Rewards \cite{guo2025deepseek}, which \tool can support through its automated verifiable bugs and intermediate steps.

\begin{table}[t]%[h!]
\centering
\small
\tabcolsep 2.5pt
\caption{Training of different model sizes on 1K trajectories on zero-shot (fine-tuned) (RQ4)}
\label{tab:train-model-size}
\vspace{-6pt}
\begin{tabular}{@{}cr|rrr@{}}
\toprule
\multicolumn{2}{c|}{Dataset} & \multicolumn{1}{c}{$\%$ Resolved $(\uparrow)$} & \multicolumn{1}{c}{$\%$ Correct $(\uparrow)$} & \multicolumn{1}{c}{$\%$ EP $(\downarrow)$} \\ \midrule
\multicolumn{1}{c|}{\multirow{3}{*}{\begin{tabular}[c]{@{}c@{}}SWE-Bench\\ Lite\end{tabular}}} & 7B & 4.67\% (8.00\%) & 0.00\% (0.67\%) & 68.67\% (33.00\%) \\
\multicolumn{1}{c|}{} & 14B & 11.00\% (11.67\%) & 1.33\% (1.33\%) & 33.33\% (25.67\%) \\
\multicolumn{1}{c|}{} & 32B & 17.33\% (17.00\%) & 1.67\% (2.00\%) & 17.33\% (15.33\%) \\ \midrule
\multicolumn{1}{c|}{\multirow{3}{*}{BugsInPy}} & 7B & 5.99\% (13.56\%) & 0.63\% (1.26\%) & 60.88\% (26.81\%) \\
\multicolumn{1}{c|}{} & 14B & 11.67\% (17.67\%) & 1.89\% (1.89\%) & 47.32\% (28.08\%) \\
\multicolumn{1}{c|}{} & 32B & 23.34\% (23.97\%) & 3.47\% (2.21\%) & 18.61\% (15.46\%) \\ \bottomrule
\end{tabular}
\end{table}

\begin{comment}

\subsubsection*{Impact of Model Size for sampling trajectories on {\tool}}

\textbf{$RQ_5$: How does model performance vary across different model sizes used for sampling trajectories when being fine-tuned on our synthetic training data?}

To access the impact of the teacher model for generating quality trajectories, we conducted an experiment where we vary the model size in this stage. We sample trajectories from \code{DeepSeek-v3-0324} \cite{deepseek-ai2024deepseek0v3} and \code{Qwen Coder 2.5 Instruct 32B}. For both models, we sample trajectories for the first round using temperature of 0 and collect trajectories that successfully solve the bug by executing corresponding unit tests. Then we sample trajectories for the remaining bugs using using temperature of 1 and adding successful trajectories to the training set. We then use those trajectories to train \code{Qwen Coder 2.5 Instruct 32B} model using deepspeed-zero-3 \cite{10.5555/3433701.3433727}.
We also use \code{Salesforce/SFR-Embedding-Code-2B\_R}
Context 32K instead
% temp 0: 1666 traj (solved 1666/9459)
% temp 1: 470 traj (solved 470/7793)
% 2136 traj
% we thus sample 2136 traj from dataset for training
% For a fair comparison, we 

Table ~\ref{tab:} reveals that ...
\end{comment}

\subsection{In-depth Study of Data Synthesis Pipeline}

\begin{table}[t]
    \centering
    \tabcolsep 2.5pt
    \small
    \setlength{\tabcolsep}{3pt}  % reduce spacing between columns
    \caption{APR performance when training on data synthesized at different granularity levels. Difficulty is defined as \% of unresolved variants by agent over total \# of variants (RQ5)}
    \label{tab:component-granularity-ablation}
    \vspace{-6pt}
\begin{tabular}{@{}c|r|crr@{}}
\toprule
\multirow{2}{*}{Level of granularity} & \multicolumn{1}{c|}{\multirow{2}{*}{\begin{tabular}[c]{@{}c@{}}Difficulty\\ $(\%, \uparrow)$\end{tabular}}} & \multicolumn{3}{c}{Trained model performance} \\ \cmidrule(l){3-5} 
 & \multicolumn{1}{c|}{} & $\%$ R $(\uparrow)$ & \multicolumn{1}{c}{$\%$ C $(\uparrow)$} & \multicolumn{1}{c}{$\%$ EP $(\downarrow)$} \\ \midrule
Class level & 94.07\% & \multicolumn{1}{r}{9.67\%} & 1.67\% & 29.00\% \\ \midrule
Function level & 87.03\% & \multicolumn{1}{r}{12.00\%} & 1.67\% & 35.67\% \\ \midrule
% \bottomrule
\end{tabular}
\end{table}

\subsubsection{Impact of component granularity}
\textbf{$RQ_5$: How do different
component granularities affect the trained models’ performance?}

We fine-tuned the model using 400 successful rollout trajectories generated by {\tool} under each granularity level.

As shown in Table \ref{tab:component-granularity-ablation}, fine-tuning on function-level granularity generated trajectories achieved the highest resolution rate of 12\%, with an empty patch rate of 35.67\%. In contrast, class-level granularity resulted in a lower resolution rate of 9.67\% but had the lowest empty patch rate at 29\%. Variants generated at the class level posed the greatest challenge for repair, with 94.07\% remaining unresolved by Qwen 2.5 Coder Instruct 32B using Moatless Tools — compared to 87.03\% for function-level granularity. The result suggests that {\em training on larger granularity does not necessarily lead to improved APR performance}. 

\subsubsection{Impact of component selection}
\label{sec:component-selection-exp}

\textbf{$RQ_6$. How do different com-
ponent selection strategies affect the trained models’ performance?}

We employ two component-selection strategies for rewriting: uniform random selection and test coverage-based selection (Section \ref{subsection:component-selection}). To evaluate their impact, we fine-tuned on 400 success-rollout trajectories generated by each strategy. As seen in Table \ref{tab:component-selection}, coverage-based selection achieves a higher resolve rate (12.33\% vs. 12.00\%) and a lower empty patch rate (26.67\% vs. 35.67\%) compared to uniform sampling. This aligns with prior observations that trivial or irreproducible bugs often lead to more empty patches~\cite{yang2024swe}. Moreover, coverage-based selection produces variants harder to fix, with 87.39\% of them remaining unresolved by Qwen 2.5 Coder Instruct 32B + Moatless Tools (versus 84.09\% for uniform selection). By focusing on components covered by more tests, the coverage-based strategy synthesizes a variety of realistic, non-trivial bugs.

\begin{table}[t]
    \centering
\small
\caption{Performance with different component selection. Here difficulty is defined as percentages of unresolved bug variants by agent over total number of variants. (RQ6)}
    \label{tab:component-selection}
    \vspace{-6pt}
\begin{tabular}{@{}c|c|crr@{}}
\toprule
\multirow{2}{*}{\begin{tabular}[c]{@{}c@{}}Component\\ selection strategy\end{tabular}} & \multirow{2}{*}{\begin{tabular}[c]{@{}c@{}}Difficulty\\ $(\%, \uparrow)$\end{tabular}} & \multicolumn{3}{c}{Trained model performance} \\ \cmidrule(l){3-5} 
 &  & $\%$ R $(\uparrow)$ & \multicolumn{1}{c}{$\%$ C $(\uparrow)$} & \multicolumn{1}{c}{$\%$ EP $(\downarrow)$} \\ \midrule
Uniform sampling & \multicolumn{1}{r|}{87.03\%} & \multicolumn{1}{r}{12.00\%} & 1.67\% & 35.67\% \\ \midrule
\begin{tabular}[c]{@{}c@{}}Test coverage\\ based sampling\end{tabular} & \multicolumn{1}{r|}{89.39\%} & \multicolumn{1}{r}{12.33\%} & 2.00\% & 26.67\% \\ \bottomrule
\end{tabular}
\end{table}

\subsubsection{Impact of model size}

\textbf{$RQ_7$: How does the size of the model used for component rewriting affect APR performance?}

For this experiment, we employed three Qwen 2.5 Coder Instruct variants—0.5B, 3B, and 14B parameters—to re-implement selected components at a temperature of 0.7. We used same number of trajectories (182) for all three models. Notably, the 0.5B model yields only 9.00\% “usable” variants (i.e., those that compile and fail at least one test case), largely due to frequent prompt-formatting errors and invalid syntax (e.g., unclosed braces), while this number or 3B and 14B is 57.28\% and 57.66\% respectively.

As seen in Table \ref{tab:model_size_mutation_ablation}, the performance metrics of the APR model are similar across different model sizes used for re-implementation. This result suggests that the size of the model used in re-writing has a small impact. However, as explained, the primary difference arises in data generation efficiency: larger models tend to yield fewer syntax errors and thus provide more valid training instances. Consequently, while smaller models can still produce some useful synthetic bugs, their high rate of unusable variants makes them less practical for large-scale synthetic data generation.

\begin{table}
\small
    \centering
    \caption{APR model performance with respect to different sizes of models used for re-implementation (RQ7)}
\label{tab:model_size_mutation_ablation}
\vspace{-6pt}
\begin{tabular}{@{}r|rrr@{}}
\toprule
\multicolumn{1}{c|}{\begin{tabular}[c]{@{}c@{}}Model\\ Size\end{tabular}} & \multicolumn{1}{c}{$\%$ Resolved $(\uparrow)$} & \multicolumn{1}{c}{$\%$ Correct $(\uparrow)$} & \multicolumn{1}{c}{$\%$ Empty Patch $(\downarrow)$} \\ \midrule
0.5B & 10.33\% & 1.33\% & 31.67\% \\
3B & 9.67\% & 1.33\% & 42.00\% \\
14B & 10.67\% & 1.33\% & 25.33\% \\ \bottomrule
\end{tabular}
\end{table}

\begin{table}
    \centering
    \small
    \caption{Fix generation strategy performance (RQ8)}
    \label{tab:reverse-patch-diff-performance}
    \vspace{-6pt}
\begin{tabular}{@{}l|rrr@{}}
\toprule
Fix generation strategy & \multicolumn{1}{l}{$\%$ R $(\uparrow)$} & \multicolumn{1}{l}{$\%$ C $(\uparrow)$} & \multicolumn{1}{l}{$\%$ EP $(\downarrow)$} \\ \midrule
RAG reverse patch diff & 0.67\% & 0.00\% & 3.00\% \\
RAG rejection sampling (w/o int.) & 3.00\% & 0.33\% & 9.00\% \\ 
Rollout zero-shot & 11.00\% & 1.33\% & 33.33\% \\ 
Rollout rejection sampling & 15.33\% & 1.33\% & 24.33\% \\ \bottomrule
\end{tabular}
\end{table}

\subsubsection{Fix ground-truth extraction strategies}
\label{sec:rq8}

\textbf{$RQ_8$: How well does the model perform when being trained on reverse patch diff compared to when being trained on {\tool} with rollout?}

We investigate three strategies for extracting training patches from synthetic bug variants:  
(1) \textbf{Naive Reverse Patch Diff}: the diff between the variant and the original component is used as the ground-truth fix (Section \ref{sec:groundtruth-extraction}), and a gold RAG model~\cite{jimenez2024swebench} is trained with the buggy file and error log.  
(2) \textbf{RAG with Rejection Sampling}: Qwen 2.5 Coder generates candidate diffs, and only those that pass tests are retained, filtering out unnatural fixes but lacking intermediate repair steps.  
(3) \textbf{Rollout Rejection Sampling ({\tool})}: patches are derived from successful rollout trajectories of repair actions (Section \ref{sec:groundtruth-extraction}).

\vspace{2pt}
{\bf Quantitative Analysis.} Table~\ref{tab:reverse-patch-diff-performance} shows that reverse patch diff yields a very low resolve rate (0.67\%) despite a low empty patch rate (3.00\%). RAG with rejection sampling improves to 3.00\% resolve rate but still underperforms the zero-shot baseline (11.00\%). Rollout rejection sampling achieves the best performance with a 15.33\% resolve rate, demonstrating the value of intermediate repair steps.

\begin{figure}
    \centering
    \includegraphics[width=1.02\linewidth]{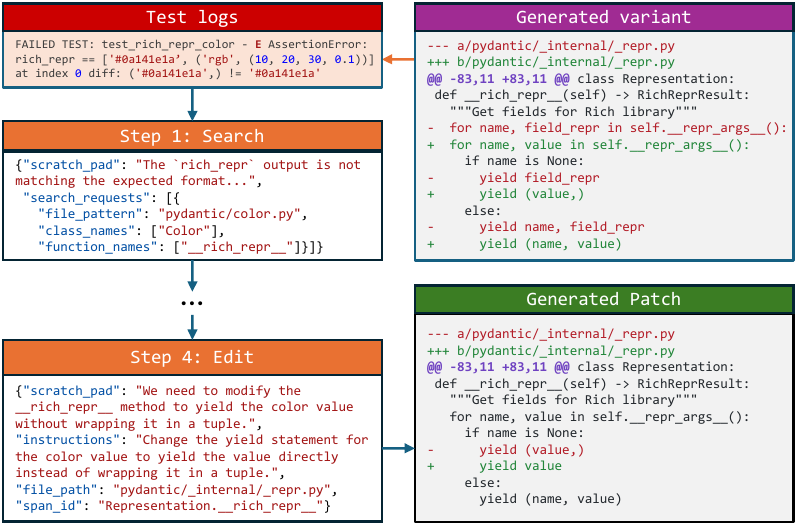}
    \vspace{-12pt}
    \caption{A generated fix with intermediate steps}
    \label{fig:intermediate-steps}
\end{figure}

{\bf Qualitative Analysis.} Figure~\ref{fig:intermediate-steps} illustrates a \code{pydantic} bug involving the \code{\_\_rich\_repr\_\_} method. Without access to reverse diffs, {\tool} leverages the Moatless agent to analyze logs and iteratively generate repair steps. As shown, it retrieves context (Step 1), reasons about modifications (Steps 2–4), and ultimately produces a correct patch distinct from the original code.

\subsection{Threats to Validity}
\label{sec:limit}

First, our synthetic bugs may not fully capture the complexity and diversity of real-world defects in specific domains. For this, one can follow our method to derive the bug dataset from real-world projects in a new domain. Second, the correctness of intermediate repair steps is automatically verified, however, they might not be representative for all alternative repair solutions. We also need to evaluate further on the usefulness of intermediate steps. Third, we used the projects in SWE-Gym as the bases for LLM reimplementation. The quality of {\tool} might be affected by such choice. Finally, our evaluation relies on existing benchmarks (SWE-Bench and BugsInPy), which, while widely used, may not comprehensive.
\section{Related Work}

\subsubsection*{Datasets for Automated Program Repair (APR)}

Training an APR model requires datasets of bug–fix pairs. These datasets typically fall into three categories. First, mined datasets from open-source repositories~\cite{muennighoff2023octopack, wei2025swerl, jimenez2024swebench, pan2024trainingsoftwareengineeringagents}, such as BugsInPy~\cite{BugsInPy}, Defects4J~\cite{just2014defects4j}, and FixEval~\cite{haque2023fixevalexecutionbasedevaluationprogram}, provide realistic data at scale. However, they often suffer from issues like tangled commits~\cite{herzig2013it}, missing test cases, and the absence of intermediate reasoning traces, limiting their effectiveness for training structured APR workflows.

Second, manually curated datasets—such as SWE-Bench~\cite{jimenez2024swebench} and SWE-Gym~\cite{pan2024trainingsoftwareengineeringagents}—offer structured patches and executable environments with tests. These enable reasoning trace extraction but are inherently hard to scale due to the extensive human effort required. Recent efforts like SWE-Smith~\cite{yang2025swesmith} have explored automated pipelines for scaling data collection for SWE agents.

Third, synthetic data generation offers a scalable alternative. Early efforts~\cite{xu2024wizardlm, luo2024wizardcoder, wei2024magicoder, yuan2024advancing} have explored using LLMs to generate synthetic data, but these are often restricted to function-level tasks and lack full repository-level semantics, test-based verifiability, or repair traces. Recent work on synthetic code datasets, such as KodCode~\cite{xu2025kodcode} for diverse coding challenges and HexaCoder~\cite{hajipour2024hexacoder} for secure code generation, demonstrates the potential of oracle-guided synthetic training. However, their application to repository-level APR tasks with executable verification remains limited. Our work focuses on creating repository-level APR datasets that are verifiable, context-rich, and scalable, addressing this gap.

\subsubsection*{Automated Program Repair and Programming Agents}

Modern learning-based APR methods often frame bug fixing as a translation task using Neural Machine Translation (NMT) models~\cite{10.1145/3551349.3556926, 10.1145/3510003.3510177, 10172686, drain2021deepdebugfixingpythonbugs}, building on advances in code understanding and representation learning~\cite{bui2019towards, nguyen2024hierarchynet, to2024functional}. While effective on simple benchmarks, they rely heavily on large-scale historical data and often suffer from generalization issues due to noise and lack of structured reasoning.

LLMs have introduced a paradigm shift, enabling zero-shot repair capabilities from natural language or code context~\cite{10.1145/3540250.3549101, 10172803}. This has inspired LLM-based agent frameworks for APR that simulate human workflows with structured tool usage and reasoning~\cite{yang2024swe, moatless_tool, antoniades2025swesearchenhancingsoftwareagents, zhang2024autocoderover, xia2024agentlessdemystifyingllmbasedsoftware, wang2024openhandsopenplatformai, phan2024hyperagent, nguyen2024agilecoder}. Notable approaches include RepairAgent~\cite{bouzenia2024repairagent}, an autonomous LLM-based agent for program repair, and ThinkRepair~\cite{yin2024thinkrepair}, which uses self-directed reasoning. RepairLLaMA~\cite{silva2025repairllama} demonstrates efficient fine-tuning strategies, while AlphaRepair~\cite{xia2024alpharepair} frames repair as code generation from bugs. Process-aware approaches like REPAIR~\cite{zhao2024repair} incorporate feedback during the repair process. However, these agents typically rely on proprietary LLMs due to the lack of scalable, high-quality training data for open models—a gap our work directly addresses.

Fault localization is a critical component of APR. Recent work has explored LLM-based fault localization~\cite{yang2024large_fault_localization}, with AgentFL~\cite{qin2024agentfl} scaling these techniques to project-level context. Our synthetic dataset generation implicitly provides fault localization signals through test failures, enabling models to learn localization alongside repair.

Efforts such as SWE-RL~\cite{wei2025swerl} and LingMa~\cite{ma2024lingmaswegptopendevelopmentprocesscentric} introduced trajectory filtering via patch similarity and fault localization to improve training quality. Still, they are prone to noisy signals and lack verified environments for evaluating correctness. Our dataset complements these efforts by offering scalable, verifiable, process-aware synthetic traces for agent training and evaluation.

\subsubsection*{Agent Training Environments}

SWE agent training requires executable and diverse environments. SWE-Bench~\cite{jimenez2023swe} provides gold patches but lacks execution scaffolding. R2E~\cite{jain2024r2e} includes 246 instances with partial environment simulation. SWE-Gym~\cite{pan2024training} introduces executable tasks with human-curated test cases, enabling structured training but limiting scalability. Our framework contributes a synthetic, verifiable, and scalable alternative. Inspired by synthetic dataset creation in other fields~\cite{long2024llms}, our mutation and trace-generation pipeline offers rich repository-level training environments for agents.

\subsubsection*{Verifiers and Test Generation}

Verifiers have become integral to improving patch quality. Execution-based verifiers~\cite{xia2024agentless} run generated patches against tests to validate correctness, while execution-free methods~\cite{pan2024training, zhang2024diversity} score agent trajectories or use LLM-based judges. Each approach has strengths: execution-based methods offer precision but struggle with ambiguity; execution-free verifiers are flexible but may be misled by surface heuristics~\cite{gu2024counterfeit}. Recent work proposes hybrid strategies to combine the best of both~\cite{zhang2023algo, ridnik2024code}.

Test generation has emerged as a complementary technique for APR verification. CodaMOSA~\cite{Lemieux2023CODAMOSA} combines search-based testing with LLMs to escape coverage plateaus, while Otter~\cite{ahmed2025otter} generates tests from issue descriptions to validate SWE patches. These approaches highlight the importance of test-based verification in APR workflows. We complement these efforts by generating data with built-in test signals and trace-level granularity, allowing verifiers to be trained and evaluated on both execution outcomes and reasoning paths.

\section{Conclusion}

In this work, we introduced \tool, a framework to overcome data scarcity in training LLMs for automated program repair. \tool employs a repository-level mutation strategy to generate large-scale, verifiable bug-fix data with human-like bugs, test cases for verification, and repair trajectories capturing the iterative debugging process.

Our contributions are threefold: (1) a scalable framework for producing vast synthetic bug data with minimal human oversight, (2) release of the \tool dataset containing bug-fix pairs, test suites, and intermediate repair steps, and (3) empirical validation showing an LLM agent trained on \tool achieves a {\em 15.33\% resolve rate} on SWE-Bench Lite, surpassing the {\em 13\%} rate of agents trained on manual datasets like SWE-Gym.

These results highlight synthetic data as a scalable, cost-effective alternative to manual collection, advancing autonomous software engineering and enabling more powerful, accessible automated bug-fixing solutions.
\vspace{2pt}
{\bf Novelty and Significance.} \underline{First}, by enabling scalable and diverse bug generation, {\tool} addresses the longstanding data scarcity challenge in APR, paving the way for more robust and generalizable repair models via intermediate steps. \underline{Second}, our methodo is generic for any programming language due to the inherent capability in LLMs. \underline{Third}, it paves ways for research on exploring adaptive bug synthesis tailored to different contexts including {\em specific programming paradigms, security vulnerabilities, or domain-specific software}, etc. \underline{Fourth}, this method can be extended to generate adversarial bugs, facilitating the evaluation of APR models under realistic conditions. \underline{Finally}, the integration of synthetic bugs can enhance automated testing and vulnerability detection by refining models to handle faults in different domains.

\bibliographystyle{plainnat}
\bibliography{main}

\end{document}